# On Polar Magnetic Field Reversal in Solar Cycles 21, 22, 23, and 24

**Mykola I. Pishkalo**

Main Astronomical Observatory, National Academy of Sciences, 27 Zabolotnogo vul., Kyiv, 03680, Ukraine

pish@mao.kiev.ua

**Abstract** The Sun's polar magnetic fields change their polarity near the maximum of sunspot activity. We analyzed the polarity reversal epochs in Solar Cycles 21 to 24. There were a triple reversal in the N-hemisphere in Solar Cycle 24 and single reversals in the rest of cases. Epochs of the polarity reversal from measurements of the *Wilcox Solar Observatory* (WSO) are compared with ones when the reversals were completed in the N- and S-hemispheres. The reversal times were compared with hemispherical sunspot activity and with the *Heliospheric Current Sheet* (HCS) tilts, too. It was found that reversals occurred at the epoch of the sunspot activity maximum in Cycles 21 and 23, and after the corresponding maxima in Cycles 22 and 24, and one-two years after maximal HCS tilts calculated in WSO. Reversals in Solar Cycles 21, 22, 23, and 24 were completed first in the N-hemisphere and then in the S-hemisphere after 0.6, 1.1, 0.7, and 0.9 years, respectively. The polarity inversion in the near-polar latitude range $\pm(55–90)°$ occurred from 0.5 to 2.0 years earlier that the times when the reversals were completed in corresponding hemisphere. Using the maximal smoothed WSO polar field as precursor we estimated that amplitude of Solar Cycle 25 will reach 116±12 in values of smoothed monthly sunspot numbers and will be comparable with the current cycle amplitude equaled to 116.4.

## 1. Introduction

Studying of evolution of the Sun's polar magnetic fields is very important for understanding and forecasting of solar activity. Polar fields change their polarity in the epoch of solar maximum and remain predominantly unipolar during the most part of solar cycle. Polar fields are open, and their polarity and distribution determine the global structure of solar corona and the interplanetary magnetic field (see, for example, review by Petrie, 2015, and references therein).

There are some difficulties in observation of the polar fields because they are observed from the Earth at large projection angle, and, moreover, the Sun's northern and southern poles are invisible during some period due to the 7.25° inclination of the Earth's orbit to the plane of the solar equator. To the same, the polar fields are in general significantly weaker than magnetic fields of sunspots and other active features on the middle and low latitudes.

Hale (1913) first reported about observed weak general magnetic field of the Sun with opposite polarity in the N- and S-hemispheres. Harold Babcock first observed the asymmetric pattern of the polar reversal. He found that in Solar Cycle 19 the polarity reversal in the S-hemisphere preceded the N-hemisphere by approximately 18 months (Babcock, 1959). The shift in time in the polarity inversion at one pole relative to the other causes an apparent situation when the both poles have the same polarity. It is, of course, only apparent, not real, monopole magnetic situation at the Sun.

The evolution of the Sun's polar magnetic fields is well consistent, in general, with Horace Babcock's phenomenological model for the solar cycle (Babcock, 1961), with Leighton's numerical kinematic flux transport model (Leighton, 1969) and, in more details, with advanced modern theories of αΩ-dynamo of the Babcock–Leighton type (see reviews by Charbonneau, 2010; Ossendrijver, 2003; Petrie *et al.*, 2014, and references therein).

The polar fields have been actively observed since the 1960s at many observatories. Now they are regularly observed from the ground at the *Kitt Peak National Observatory* and at the *Wilcox Solar Observatory* (WSO), and from space using *Helioseismic and Magnetic Imager* onboard *Solar Dynamics Observatory* and *Solar Optical Telescope* onboard the *Hinode* spacecraft.

As a rule, in the N- and S-hemispheres, the polar field changes its polarity non-simultaneously. The asymmetry, or asynchrony, in the reversal process reflects the solar activity asymmetry between the hemispheres (Svalgaard and Kamide, 2013), especially asymmetry of the high-latitudinal magnetic fields (Mordvinov and Yazev, 2014).

The reversal of the averaged magnetic field at high latitudes, over $\pm 55°$ latitude, occurs approximately a year earlier than directly at the poles (Pishkalo *et al.*, 2005).

Polar field distribution can be studied by indirect methods too, such as distribution of filaments or coronal holes. Filaments (prominences) at high latitudes mark neutral lines between regions with predominantly unipolar and opposite-polarity magnetic flux. Coronal holes, in turn, mark regions with unipolar open magnetic flux.

Using the Hα synoptic charts and the data on the polar prominences, Makarov and coauthors (Makarov *et al.*, 1983; Makarov and Sivaraman, 1983; Makarov and Sivaraman, 1986) studied filaments and filament channels distribution at high latitudes in Solar Cycles 11–21 and found epochs of polarity reversals. Moreover, they reported that the reversals repeatedly occur at the poles, *i.e.* triple reversals were found in the N-hemisphere in Cycles 16, 19, and 20 and in the S-hemisphere in Cycles 12 and 14.

In Solar Cycles 21–23 single reversals were identified in both hemispheres according to former results (Durrant and Wilson, 2003; Makarov and Sivaraman, 1986; Pishkalo *et al.*, 2005; Snodgrass *et al.*, 2000; Webb *et al.*, 1984). On the other hand, Mordvinov and Yazev (2014) concluded that a triple reversal occurred at the north in Cycle 21.

Many papers are devoted to the polar field reversals in the current Cycle 24. However, the reversal times for Cycle 24 reported by different researchers were different. In particular, Shiota *et al.* (2012) concluded that the reversal in the N-hemisphere should finish to the middle of 2012. Svalgaard and Kamide (2013) pointed out that, as expected, the reversal first took place in the N-hemisphere. Karna *et al.* (2014) found that the N-pole changed its polarity in the middle of 2012 and the S-hemisphere will go through reversal at least after 1.5 years later. Upton and Hathaway (2014) forecasted the reversal of the global dipole field of the Sun to be completed by December 2013.

Mordvinov and Yazev (2014) found that in Solar Cycle 24 in the N-hemisphere the reversal ended to rotation 2036 (in early 2013). Sun *et al.* (2015) obtained that, at latitudes over $\pm 60°$, the averaged magnetic field changed its polarity in the N- hemisphere in November 2012 and in the S-hemisphere in March 2014, respectively.

Investigation of Pastor Yabar *et al.* (2015) indicated that the reversals in the N- and S-hemisphere were completed in April 2014 and February 2013, respectively.

Pishkalo and Leiko (2016) reported that triple reversal in the N-hemisphere and single reversal in the S-hemisphere were completed to the middle of 2014 (at the beginning of rotation 2150) and in April–May 2015 (at the beginning of rotation 2162), respectively.

Gopalswamy *et al.* (2016) found that the reversal in the S-hemisphere occurred around June 2014 and in the N-hemisphere only by October 2015.

Janardhan *et al.* (2018) found that the reversal in the N-hemisphere was multiple and completed only by November 2014, and in the S-hemisphere it was single and clean reversal, which finished in November 2013.

Polar field in the cycle minimum can be considered as a physically-based precursor for solar activity in the maximum of solar cycle (Schatten *et al.*, 1978). There are several predictions of maximal amplitude of Solar Cycle 25, to date, based on the polar field evolution before the cycle minimum (Cameron *et al.* 2016; Upton and Hathaway, 2018; Wang, 2017).

The aim of this work is to compare the time of polarity reversal in the latitude range ±(55–90)° with the time when reversals was finished at the poles, in Solar Cycles 21 to 24, and to estimate the maximal amplitude of the next Cycle 25 using measurements of polar magnetic fields before the cycle minimum as a precursor.

**2. Data**

For the analysis we used monthly and monthly smoothed international sunspot numbers (since 1975) and the corresponding hemispherical values (since 1992) from the *Sunspot Index and Long-term Solar Observations* (SILSO, http://sidc.oma.be/SISLO, Version 2.0). Hemispherical sunspot numbers from 1975 to 2000, published by Temmer *et al.* (2002), were also used to analyze period before 1992.

Polar magnetic fields measured at the WSO (http://wso.stanford.edu) and the *Heliospheric Current Sheet* (HCS) tilts calculated by the WSO team in the line-of-sight and radial approaches of the PFSS-model were also analyzed. It should be noted that the polar field strength values, which were determined at the WSO and are used in this paper, are not strength of magnetic field at the poles directly. They represent averaged magnetic fluxes from about the ±55° latitude to the pole for the N- and S-hemisphere. The HCS tilts are maximal latitudes of the magnetic neutral lines at the source surface calculated for each Carrington rotation; they are limited at about ±75° latitude.

It should also be mentioned that usually the strength of polar fields are not measured at poles directly. The narrower is the latitude range near the pole where the strength of polar field is measured (or averaged), the closer the time of polarity inversion obtained to the real time when the reversal process is completed in corresponding hemisphere.

Epochs of polar magnetic field reversal, *i.e.* times when the reversal process was completed in each hemisphere, in Solar Cycles 21, 22, and 23 were taken from the paper (Pishkalo *et al.*, 2005), where they were found from analysis of polar prominences and filaments (as a time when polar filaments/prominences disappeared at poles).

For Solar Cycle 24 we used the times when polarity reversals were completed in both N- and S-hemispheres from our paper (Pishkalo and Leiko, 2016) where averaged magnetic fields were studied in some latitudinal ranges (zones) near the poles.

Although timings of reversal for Cycles 21–23 and for Cycle 24 were obtained using different methods they can be reasonably used in our study. It was shown in many investigations that position and evolution of magnetic neutral lines inferred from magnetic observations and from Hα synoptic charts (filaments and filament/prominence bands) are in good (one-two solar rotations) agreement (see, *e.q.*, Duvall *et al.*, 1977; Makarov and Sivaraman, 1986; Durrant, 2002; Snodgrass *et al.*, 2003). Timings of reversal for the same solar cycle obtained by these methods differ usually by one-two rotations (Pishkalo *et al.*, 2005).

**3. Results**

The evolutions of international sunspot number and the modulus of the WSO polar field strength, from 1975 to the present time, are shown in Figure 1. The smoothed values are represented by solid lines.

One can see that solar activity is gradually diminished from Cycle 21 to Cycle 24. Solar Cycles 22, 23, and 24 are clearly two-peaked. The maximal smoothed sunspot number in Cycle 24 is only about a half of corresponding value in Cycle 21.

Polar magnetic fields are maximal and minimal in the epoch of the activity minimum and maximum, respectively. The strength of polar field, measured before minima of Cycles 24 and 25, are similar. It equals to only a half of the magnitude of polar magnetic field at the minimum of Solar Cycle 22.

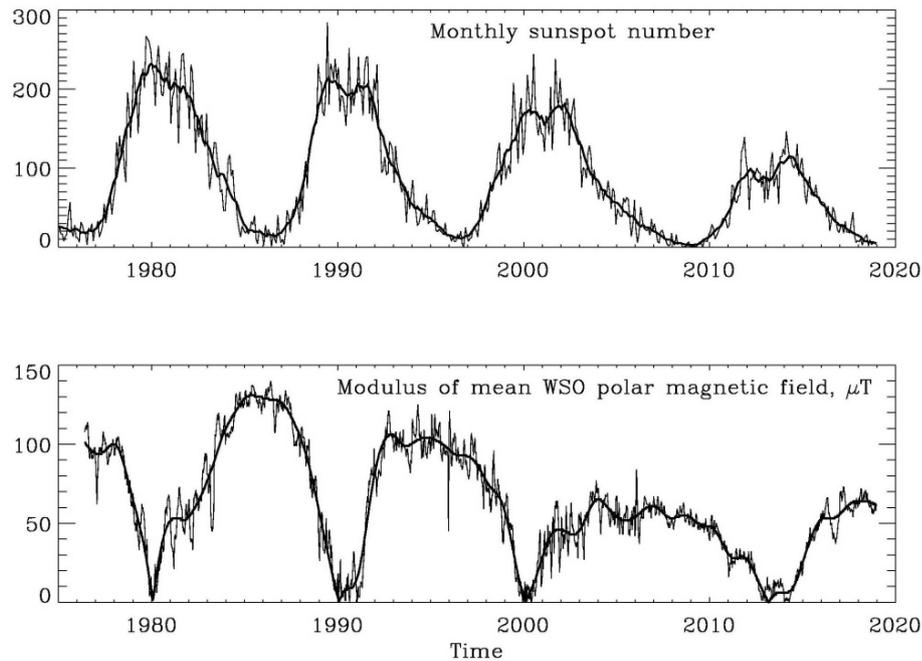

**Figure 1** Monthly sunspot number (*top*) and modulus of the mean polar field measured at the WSO (*bottom*) with time in Solar Cycles from 21 to 24. Smoothed values are shown by *thick lines*.

Time evolution of smoothed polar magnetic fields, observed at WSO in Solar Cycles from 21 to 24 in the latitude range from ±55° to poles, is shown in Figure 2. Polar fields observed in the N- and S-hemispheres are shown by solid and dotted lines, respectively. Epochs of polar field reversals in the latitude range ±(55–90)° and at poles (when the reversals were completed) are indicated by thin and thick vertical lines.

Figure 3 illustrates the HCS tilts in both hemispheres in time in Solar Cycles from 21 to 24. Epochs of polar field reversals are shown by vertical lines. Classic (line-of-sight) and radial HCS tilts are plotted here by thin and thick solid lines, respectively.

Sunspot activity in the N- and S-hemispheres separately, since 1975, and epochs of polar magnetic field reversals are shown in Figure 4. Data from Temmer *et al.* (2002) for years 1975–2000 and from SILSO for years 1992–2018 are presented by thin and thick lines, respectively. Taking into account revision of international sunspot numbers acting since July of 2015 (Clette *et al.*, 2014) and the calculated ratio of "new" and "old" sunspot numbers (Pishkalo, 2016), data by Temmer *et al.* (2002) were multiplied by 1.5 to compare them with the SILSO data presented in the "new" revised scale of sunspot numbers.

3.1. Cycle 21

Northern and southern poles have positive and negative magnetic polarity at the beginning of the cycle, respectively. The polarity inversion occurred first in the N-hemisphere, in the end of 1979 in the latitude range 55–90°, and it was completed in the beginning of 1981. In the S-hemisphere, reversal finished in July of 1981. Reversals were completed more than a year after maximal HCS tilts. Both reversals, in the latitude range ±(55–90)° and at poles, occurred at local maxima of sunspot activity in each hemisphere.

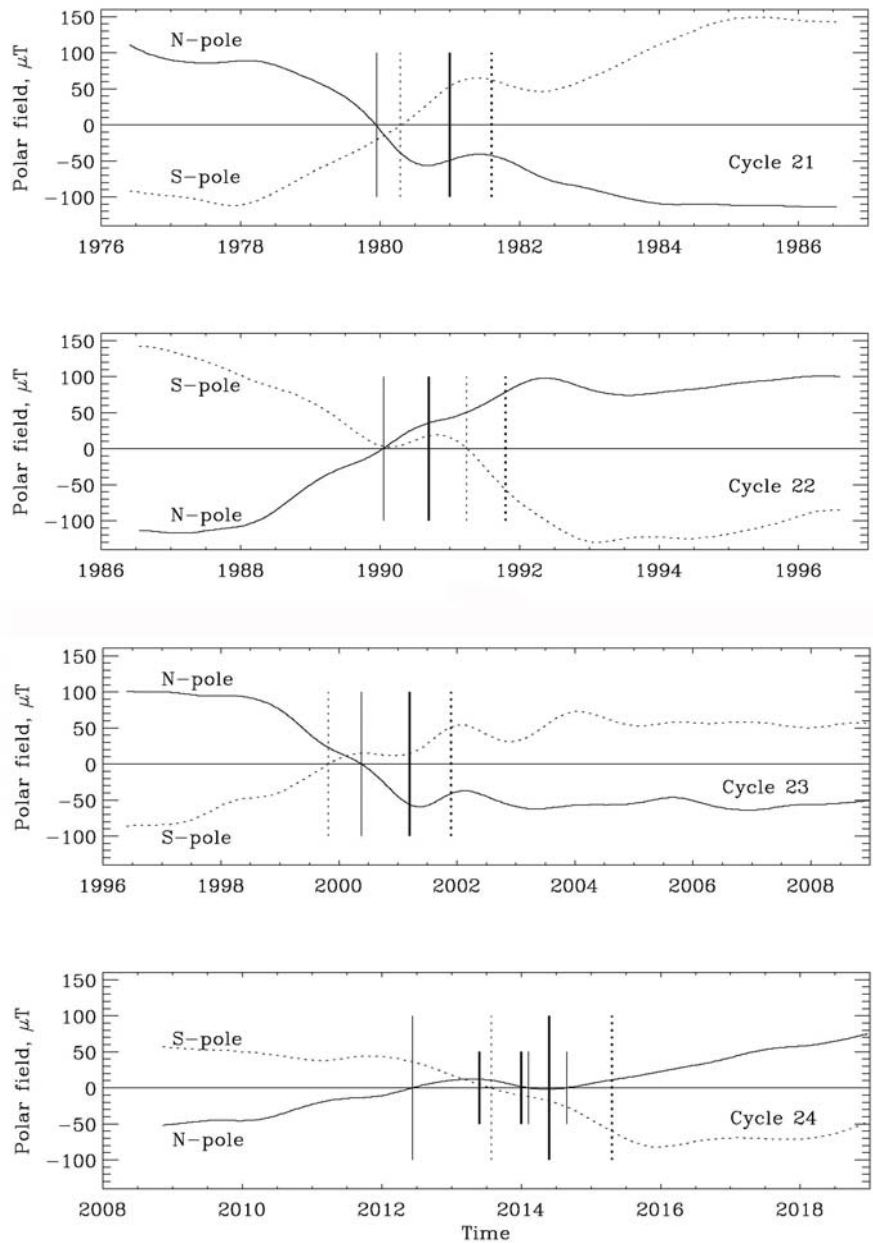

**Figure 2** Smoothed WSO polar magnetic fields in Solar Cycles 21 to 24 (from *top* to *bottom*). Times of the reversals in the latitude range ±(55–90)° and at poles are shown by thin and thick vertical lines. *Solid and dotted lines* represent N- and S-hemispheres. *Short vertical lines*, *thin* and *thick*, at *bottom* panel (Cycle 24) correspond to second and third reversals in the latitude range 55–90° and to first and second reversals at the N-pole, respectively.

### 3.2. Cycle 22

Polar magnetic field changes its polarity first in the N-hemisphere. In the S-hemisphere polar magnetic field measured by WSO became very small in the beginning of 1990, did not change its sign then finally changed polarity in about 1991.2. Reversal in the S-hemisphere was completed more than a year after the polarity inversion in the N-hemisphere. The reversal in the N-hemisphere occurred at maximum of the HCS tilts, but after maximum of hemispheric sunspot number. The reversal in the S-hemisphere, vice versa, occurred after maximum of the HCS tilts, but near the main maximum of hemispheric sunspot number.

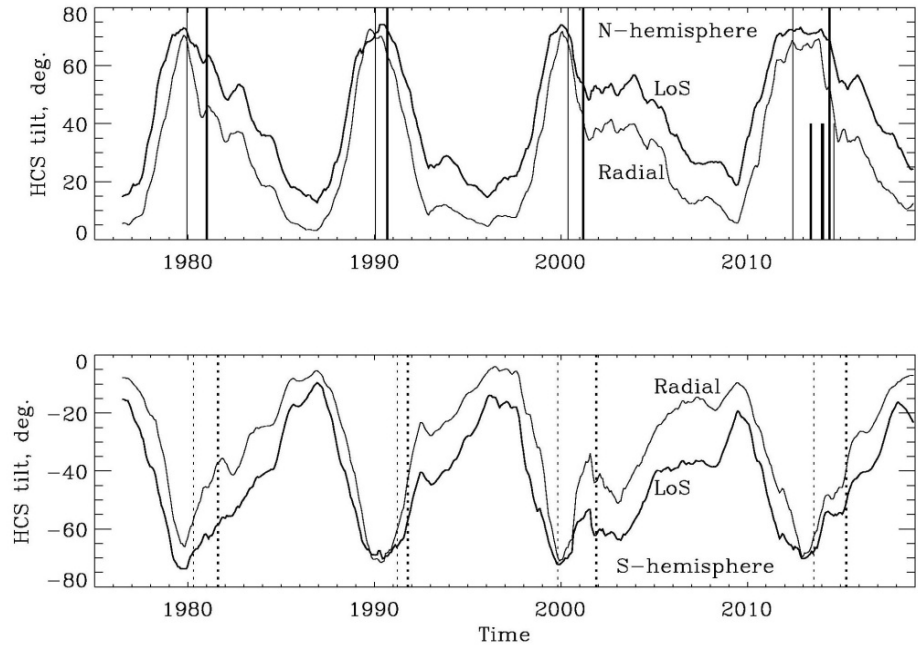

**Figure 3** The HCS tilts calculated by WSO in radial (*Radial*) and classic line-of-sight (*LoS*) approach with time in Solar Cycles from 21 to 24 (*top* and *bottom* panels are for N- and S-hemispheres, respectively). *Vertical lines* represent time of polar field reversals (see caption for Figure 2). *Note:* The HCS tilt in the N(S)-hemisphere is the maximal N(S)-latitude of magnetic neutral line at the source surface, calculated in the PFSS-model, for each solar rotation.

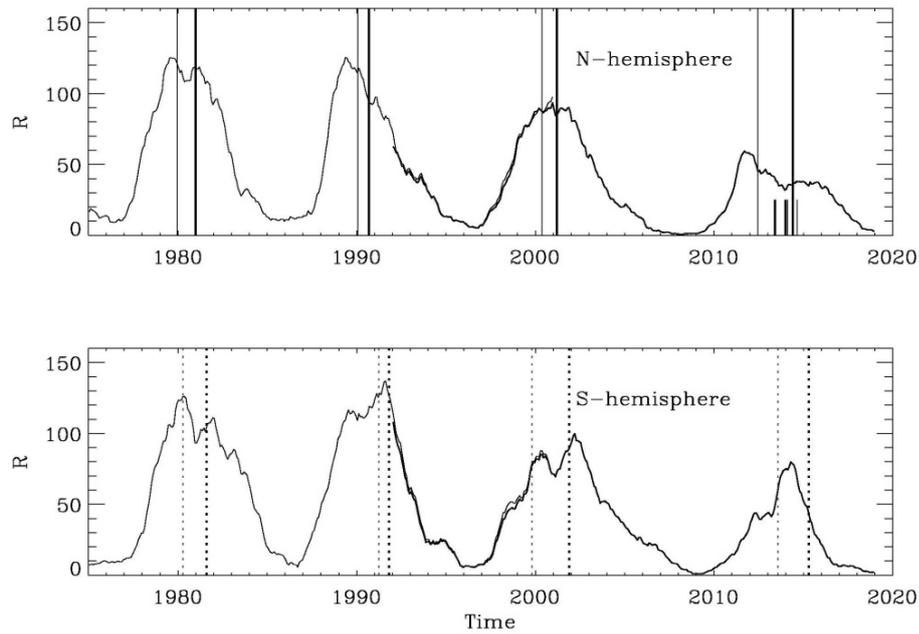

**Figure 4** Smoothed sunspot number $R$ for N-hemisphere (*top*) and for S-hemisphere (*bottom*) with time. *Thin* and *thick* lines correspond to data from Temmer *et al.* (2002) for years 1975–2000 and from SILSO for years 1992–2018. *Vertical lines* represent time of polar field reversals (see caption for Figure 2).

3.3. Cycle 23

The polarity inversion in the S-hemisphere started before reversal in the N-hemisphere and completed after it. Reversals occurred near local hemispherical maxima of sunspot activity. In both hemispheres reversals started near maxima of the HCS tilts. Reversals in the N- and S-hemisphere were completed to epochs of 2001.2 and 2001.9, respectively.

### 3.4. Cycle 24

Reversal in the N-hemisphere started first and completed first too. Polar field in the N-hemisphere changes its polarity three times, *i.e.* the triple reversal occurred. It is clearly seen from Figure 5, where shown is evolution of polar magnetic fields measured at WSO in the latitude range ±(55–90)° and, for comparison, calculated in four circumpolar caps using data from the SOLIS project (see, in more details, Pishkalo and Leiko, 2016). The more narrow near-polar latitude range where polar magnetic fields are averaged corresponds to the later polarity inversion. In the N-hemisphere the triple reversal was completed approximately a year earlier than single change of the polarity in the S-hemisphere (2014.4 and 2015.3).

The reversals occurred at maximum of HCS tilt in the N-hemisphere and after the corresponding maximum in the S-hemisphere. Epochs of reversal do not coincide with maximal hemispheric solar activity.

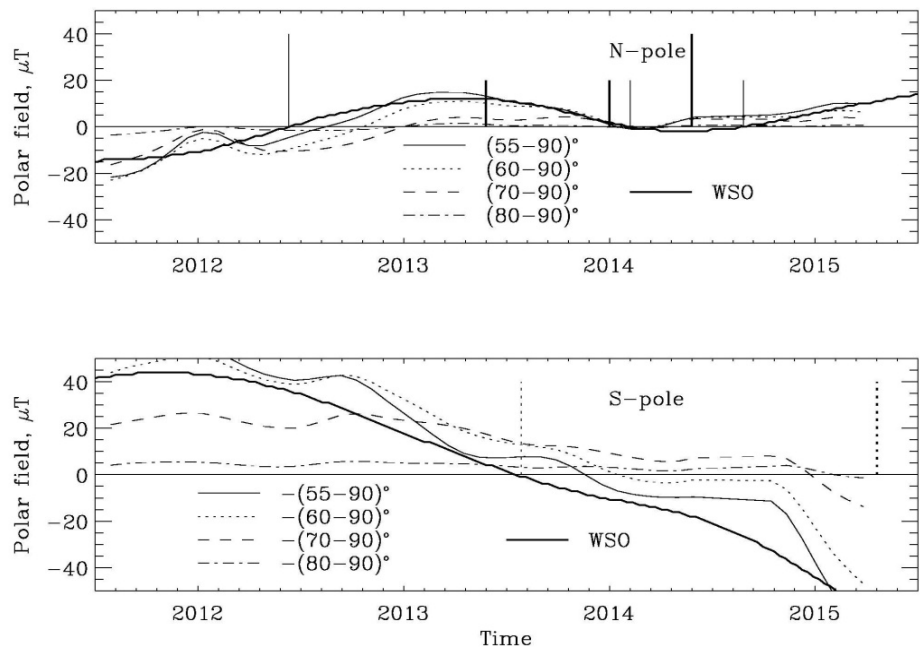

**Figure 5** Polar fields in Cycle 24 near the time of the reversals in N-hemisphere (*top*) and S-hemisphere (*bottom*). *Thick lines* represent measurements of WSO in the latitude range ±(55–90)°. Averaged polar magnetic fields calculated in several latitudinal ranges near the poles (Pishkalo and Leiko, 2016), using data from *VSM/Kitt Peak* (the SOLIS Project), are shown by *thin lines* and scaled to the WSO data. *Vertical lines* represent time of polar field reversals (see caption for Figure 2).

Information on epochs of the polarity inversion in both hemispheres in the latitude range ±(55–90)° and at poles, in Solar Cycles 21, 22, 23, and 24, is summarized in Table 1.

### 3.5. Strength of the next cycle

Polar magnetic fields observed near the cycle minimum can be a precursor for prediction of maximal sunspot activity in the cycle (Schatten *et al.*, 1978). The stronger polar magnetic fields near the cycle minimum, the higher solar cycle (see Figure 1). Polar magnetic fields near minima of Solar Cycles 24 and 25 are about of the same strength. So, we can expect that the maximal amplitude of the next Solar Cycle 25 will be approximately equaled to the amplitude of the present Cycle 24.

Maximal strength of WSO filtered polar magnetic field is observed before the cycle minimum, not exactly in the minimum. It can help us to do a simple prediction of strength of Solar Cycle 25.

Figure 6 shows relation of maximal smoothed monthly sunspot number and maximal WSO smoothed polar magnetic field. There are only three solar cycles with known corresponding parameters. Due to this reason we cannot do any serious statistical studying, but we can do estimation for the next Cycle 25. Data for Solar Cycles 22, 23, and 24 are in good agreement with the straight line connecting the "24" and "22" points and can be presented by the equation $Y = 1.466*X + 22.141$. The maximal values of WSO smoothed polar fields which were measured before the minima of Cycles 24 and 25 are equal to 65 µT and 64 µT, respectively. Taking this fact into account we can predict maximal amplitude of the next Cycle 25: 116.0±12.1 in values of smoothed monthly sunspot number.

**Table 1** Epochs of the Sun's polar magnetic field reversals in Cycles 21 to 24.

|  | Cycle 21 | Cycle 22 | Cycle 23 | Cycle 24 |
|---|---|---|---|---|
| Time of N-pole field reversal in the latitude range 55–90°, $T_{N1}$ | 1979.95 | 1990.05 | 2000.38 | 2012.44 |
| Time when N-pole reversal was completed, $T_{N2}$ | 1981.0 | 1990.7 | 2001.2 | 2014.4 |
| $\Delta T_N = T_{N1} - T_{N2}$, yrs | 1.05 | 0.65 | 0.82 | 1.96 |
| Time of S-pole field reversal in the latitude range –(55–90)°, $T_{S1}$ | 1980.29 | 1991.24 | 1999.82 | 2013.54 |
| Time when S-pole reversal was completed, $T_{S2}$ | 1981.6 | 1991.8 | 2001.9 | 2015.3 |
| $\Delta T_S = T_{S1} - T_{S2}$, yrs | 1.31 | 0.56 | 2.08 | 1.76 |
| $\Delta T = T_{S2} - T_{N2}$, yrs | 0.6 | 1.1 | 0.7 | 0.9 |

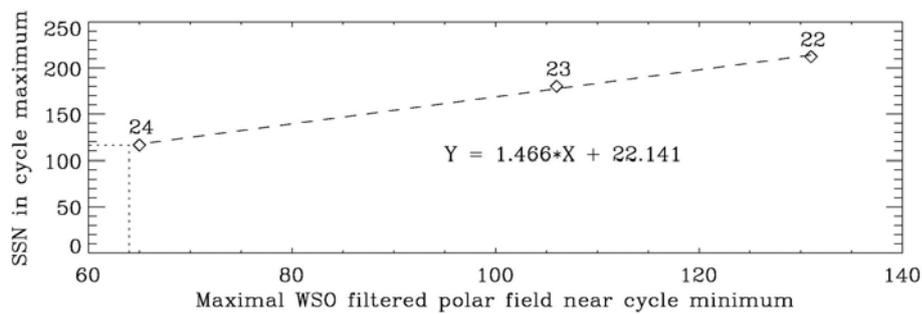

**Figure 6** Monthly smoothed sunspot number in cycle maximum (Cycles 22, 23, and 24) versus maximal WSO smoothed polar magnetic field (µT) in the latitudinal range ±(55–90)° near the cycle minimum. The best linear fit is plotted by *dashed line*. Predicted maximal smoothed monthly sunspot number for Cycle 25 is shown by *dotted line*.

## 4. Discussion and conclusion

Solar activity decreases gradually from Cycle 21 to Cycle 24 which is the weakest solar cycle in the last 100 years, after Solar Cycle 14. What will this continue further? Will solar activity decrease in the next Cycle 25 resulting in new Dalton or Maunder minimum? This question is very important now.

It is well known that variations of solar activity cause changes in the interplanetary and near-Earth space, which, in turn, affect many ground-based and space-borne technological systems (such as high-frequency radio communication and radars, ground power lines and the pipelines, space navigation and aero-navigation, etc.) and, in a certain way, also the climate and life on Earth.

Using maximal value of filtered WSO polar field strength before the cycle minimum as precursor we estimated the maximal amplitude of the next Cycle 25 to be similar to that of the present cycle. Our predicted amplitude of Solar Cycle 25 is in good agreement with other published predictions.

In particular, Wang (2017) predicted the Cycle 25 to be similar in amplitude to Cycle 24. His prediction was based on the observed evolution of polar fields and the axial dipole component of the Sun's global field as of the end of 2015.

Upton and Hathaway (2018), using their advective flux transport model, estimated that the axial dipole strength at the start of 2020 will reach to $+1.56 \pm 0.05$ G. Then they predicted that Solar Cycle 25 will be a small one, which amplitude will reach 95–97% of the amplitude of Cycle 24. This gives that Solar Cycle 25 will be the smallest one in the amplitude in the last hundred years.

Using another similar surface flux transport simulations for the descending phase of solar cycle 24 Cameron *et al.* (2016) obtained that the next Cycle 25 will be of moderate amplitude, slightly higher than in the current cycle.

Bhowmik and Nandy (2018), using magnetic field evolution models for the Sun's surface and interior, predicted that Solar Cycle 25 will be similar or slightly stronger than the current cycle and will reach its peak around 2024. On the other hand, Helal and Galal (2012) from the statistics of spotless events and using preliminarily predicted parameters of Cycle 24 found that maximal sunspot number in Cycle 25 will be 118.2, in units before the revision of sunspot numbers in 2015 (or about 177 in new units), *i.e.* more than 1.5 times higher than in Cycle 24. Similar amplitude of about 167 for Cycle 25 was predicted also in (Pishkalo, 2016) using regression equations obtained from correlations between the "new" cycle parameters. Pesnell and Schatten (2018), on the basis of the SODA index, have predicted that amplitude of Cycle 25 will equal $135 \pm 25$. According to Okoh *et al.* (2018), who used Ap index as precursor, predicted amplitude of Cycle 25 is $122.1 \pm 18.2$. These five works forecast an increase of solar activity from the next cycle.

On the contrary, recently Covas *et al.* (2019) obtained that the next Cycle 25 will have maximal amplitude of $57 \pm 17$. Their prediction, based on the so-called feed-forward artificial neural networks, indicates that Solar Cycle 25 will be one of the weakest in recorded history and that the tendency to further diminish of solar activity will remain. Javaraiah (2017) also obtained that Cycle 25 will be weaker than Cycle 24.

Janardhan *et al.* (2018) studied the process of polar field reversal in Solar Cycles 21–24 and found the unusual nature and the significant hemispheric asymmetry of the field reversal pattern in Cycle 24. They called the reversal in the current cycle "unusual", mainly, because the reversal in the N-hemisphere was triple. They found that it started in June of 2012, was followed by a long period of near-zero field strength and finished in November of 2014.

We found that the reversal in the N-hemisphere was completed earlier, in the middle of 2014. Moreover, a multiple reversal is not unique property of Solar Cycle 24. There were triple reversals in Cycles 16, 19, 20, and, perhaps, 21 in the N-hemisphere and in Cycles 12 and 14 in the S-hemisphere. Besides, there were long periods of small values of polar field strength in the S-hemisphere in Cycles 22 and 23 although they did not result in a multiple reversals (see Figure 2).

Triple, or, in common, multiple reversals seem to take place due to appearance of active regions whose polarity is opposite to the usual (according to the Hale's law) hemispheric polarity in the cycle and their further migration to the pole (Mordvinov and Kitchatinov, 2019). Violations of the evolution of polar fields, including short-term polarity reversals, can also be caused by poleward transport of magnetic flux from active regions with negative (non-Joy's) tilts. Ordinary reversals are regular attributes of the solar cycle, when magnetic flux from dispersing active regions migrate to the poles and cancel there, resulting in the polarity inversion at high latitudes (Mordvinov and Yazev, 2014; Petrie, 2015).

It seems that time lags between polar field reversals in the latitude range ±(55–90)° and at the poles (when the reversal is completed in the corresponding hemisphere) reflect speeds of meridional circulation at high latitudes. One can conclude from Table 1 that the stronger the solar cycle, the faster the meridional flow at near-pole latitudes. For example, amplitudes of Cycles 22 and 24 are 212.5 and 116.4, respectively, *i.e.* their ratio is less than 2. But the above-mentioned time lags of Cycles 22 and 24 are approximately 0.6 and 1.9 years, respectively, *i.e.* their ratio is about 1/3. Further investigations can confirm such tendency.

Results of this work can be summarized as follows:

- Reversals, or polarity inversions, of the Sun's polar magnetic fields in Solar Cycles 21 to 24 occurred near the epoch of the activity maximum in the N- and S-hemisphere, but sometimes they did not coincide with the corresponding hemispheric maxima of sunspot numbers.
- There was a triple reversal in the N-hemisphere in Solar Cycle 24. The rest of reversals were single.
- Reversals occurred approximately one–two years after maximal HCS tilts calculated at WSO.
- The polarity inversion of polar magnetic fields measured at WSO in the near-polar latitude range ±(55–90)° occurred 0.5–2.0 years earlier that the time when the reversals were completed in the corresponding hemisphere.
- Reversals of polar magnetic field in Solar Cycles 21, 22, 23, and 24 were completed first in the N-hemisphere, while in the S-hemisphere 0.6, 1.1, 0.7, and 0.9 years later, respectively.
- Maximal smoothed WSO polar field strength, used as precursor, indicates that maximal amplitude of the next Cycle 25 will be similar to the amplitude of the current cycle and will reach 116±12 in values of smoothed monthly sunspot numbers.

**Acknowledgements** The author thanks the anonymous referee for constructive comments, and the staff of *the John M. Wilcox Solar Observatory* at Stanford University and Dr. Todd Hoeksema for data on the strength of the Sun's polar magnetic field and the HCS tilts, and also the SILSO team (*Royal Observatory of Belgium*, Brussels) for data on the international sunspot numbers.

**Disclosure of Potential Conflicts of Interest**

The author declares to have no conflicts of interest.